\documentclass[journal]{IEEEtran}
\usepackage[utf8]{inputenc}

\usepackage{cite}
\usepackage[hidelinks]{hyperref}
\usepackage{footmisc}

\usepackage{bm}
\usepackage{mathtools}
\usepackage{amssymb}
\usepackage{microtype}

\usepackage[usenames,dvipsnames, table]{xcolor}
\definecolor{rot}{HTML}{D7191C}
\definecolor{blau}{HTML}{2B83BA}
\definecolor{grun}{HTML}{50B180}
\definecolor{gelb}{HTML}{FFC540}
\definecolor{lila}{HTML}{9A62C4}

\usepackage{caption}
\usepackage{subcaption}
\usepackage{adjustbox}

\usepackage{tikz}
\usetikzlibrary{intersections}
\usetikzlibrary{calc, shapes.misc, matrix}

\usepackage{pgfplots}
\usepgfplotslibrary{groupplots}
\pgfplotsset{compat=1.15}
\tikzset{cross/.style={cross out, draw=black, minimum size=2*(#1-\pgflinewidth), inner sep=0pt, outer sep=0pt, line width=1pt},
cross/.default={1pt}}

\usepackage{booktabs, dcolumn}

\usepackage[nolist]{acronym}

\usepackage{fancyhdr}
\fancypagestyle{copyright}{
        \fancyhf{}
        
        \fancyfoot[C]{\footnotesize This is the accepted version of the publiction. \textcopyright 2021 IEEE.  Personal use of this material is permitted.  Permission from IEEE must be obtained for all other uses, in any current or future media, including reprinting/republishing this material for advertising or promotional purposes, creating new collective works, for resale or redistribution to servers or lists, or reuse of any copyrighted component of this work in other works.}
}

\begin{acronym}
\acro{DFT}{discrete Fourier transform}
\acro{MVDR}{minimum variance distortionless response}
\acro{PDF}{probability density function}
\acro{MMSE}{minimum mean square error}
\acro{ML}{maximum likelihood}
\acro{SNR}{signal-to-noise ratio}
\acro{MAP}{maximum a posteriori}
\acro{ASR}{automatic speech recognition}
\acro{POLQA}{perceptual objective listening quality analysis}
\acro{MOS}{mean opinion score}
\acro{PESQ}{perceptual evaluation of speech quality}
\acro{EM}{expectation maximization}
\acro{STFT}{short-time Fourier transform}
\acro{PSD}{power spectral density}
\acro{SI-SDR}{scale-invariant signal-to-distortion ratio}
\acro{SI-SIR}{scale-invariant signal-to-interference ratio}
\acro{SI-SAR}{scale-invariant signal-to-artifacts ratio}
\acro{DNN}{deep neural network}
\acrodefplural{DNN}{deep neural networks}
\end{acronym}

\def\vec#1{\ensuremath{\bm{\mathbf{#1}}}}

\DeclareMathOperator*{\argmax}{arg~max~} 
 
\DeclareMathOperator*{\Real}{Re}

\newcommand{\ignore}[1]{}
\newcommand{\brandY}{\vec{Y}}
\newcommand{\brealY}{\vec{y}}
\newcommand{\randY}{{Y}}

\newcommand{\brandX}{\vec{X}}

\newcommand{\brandN}{\vec{N}}

\newcommand{\randN}{N}

\newcommand{\brandS}{\vec{S}}

\newcommand{\randS}{{S}}
\newcommand{\realS}{{s}}

\newcommand{\randZ}{{Z}}
\newcommand{\realZ}{{z}}

\newcommand{\realA}{{a}}

\newcommand{\covX}{\vec{C}_{x}}
\newcommand{\invCovX}{\vec{C}_{x}^{-1}}

\newcommand{\cor}{\Phi}

\newcommand{\corN}{\vec{\cor}_{n}}
\newcommand{\corM}{\vec{\cor}_{m}}

\newcommand{\invCorN}{\vec{\cor}_{n}^{-1}}
\newcommand{\invCorM}{\vec{\cor}_{m}^{-1}}

\newcommand{\psdS}{\sigma_{s}^{2}}

\newcommand{\varM}{\sigma_m^2}
\newcommand{\invVarM}{\sigma_m^{-2}}

\newcommand{\steeringvec}{\vec{d}}
\newcommand{\numMics}{\ensuremath{D}}
\newcommand{\micidx}{\ensuremath{\ell}}

\newcommand{\freqbinidx}{\ensuremath{k}}
\newcommand{\timeframeidx}{\ensuremath{i}}

\newcommand{\T}[1]{T\textsubscript{#1}}

\newcommand{\confhypergeom}{\mathcal{M}}

\newcommand{\complexGaussian}{\mathcal{CN}}

\title{Nonlinear Spatial Filtering in Multichannel Speech Enhancement}

\author{Kristina~Tesch,~\IEEEmembership{Student Member,~IEEE}, and
        Timo~Gerkmann,~\IEEEmembership{Senior Member,~IEEE}%
\thanks{The authors are with the Signal Processing Group, Department of Informatics, Universität Hamburg, 22527 Hamburg, Germany (e-mail: kristina.tesch@uni-hamburg.de; timo.gerkmann@uni-hamburg.de).}%
}

\begin{document}
\maketitle
\thispagestyle{copyright}

\begin{abstract}
The majority of multichannel speech enhancement algorithms are two-step procedures that first apply a linear spatial filter, a so-called beamformer, and combine it with a single-channel approach for postprocessing. However, the serial concatenation of a linear spatial filter and a postfilter is not generally optimal in the \ac{MMSE} sense for noise distributions other than a Gaussian distribution. Rather, the \ac{MMSE} optimal filter is a \emph{joint spatial and spectral nonlinear} function. While estimating the parameters of such a filter with traditional methods is challenging, modern neural networks may provide an efficient way to learn the nonlinear function directly from data. To see if further research in this direction is worthwhile, in this work we examine the potential performance benefit of replacing the common two-step procedure with a joint spatial and spectral nonlinear filter. 

We analyze three different forms of non-Gaussianity: First, we evaluate on super-Gaussian noise with a high kurtosis. Second, we evaluate on inhomogeneous noise fields created by five interfering sources using two microphones, and third, we evaluate on real-world recordings from the CHiME3 database. In all scenarios, considerable improvements may be obtained. Most prominently, our analyses show that a nonlinear spatial filter uses the available spatial information more effectively than a linear spatial filter as it is capable of suppressing more than $D-1$ directional interfering sources with a $D$-dimensional microphone array without spatial adaptation.

\end{abstract}


\section{Introduction}
\IEEEPARstart{I}{n} our everyday life, we are surrounded by background noise for example traffic noise or competing speakers. Hence, speech signals that are recorded in real environments are often corrupted by noise. Speech enhancement algorithms are employed to recover the target signal from a noisy recording. This is done by suppressing the background noise or reducing other unwanted effects such as reverberation. This way, speech enhancement algorithms aim to improve speech quality and intelligibility. Their fields of application are manifold and range from assisted listening devices to telecommunication all the way to \ac{ASR} front-ends \cite{doclo2015MultichannelSignalEnhancement, weninger2015speechenhancementLSTM}. 

If the noisy speech signal is captured by a microphone array instead of just a single microphone, then not only tempo-spectral properties can be used to extract the target signal but also spatial information. Spatial filtering aims at suppressing signal components from other than the target direction. The filter-and-sum beamforming approach \cite[Sec. 12.4.2]{vary2006digital} achieves this by filtering the individual microphone signals and adding them. In the frequency domain, this means to compute the scalar product between a complex weight vector and the vector of spectral representations of the multichannel noisy signal. Hence, the beamforming operation is linear with respect to the noisy input. 

The beamforming weights are chosen to optimize some performance measure. For example, minimizing the noise variance subject to a distortionless constraint leads to the well-known \ac{MVDR} beamformer \cite[Sec. 3.6]{benesty2008MicrophoneArraySignal}. The noise suppression capability of such a spatial filter alone is often not sufficient and a single-channel filter is applied to the output of the spatial filter to improve the speech enhancement performance. The second processing stage in this two-step processing scheme is often referred to as the postfiltering step.

Single-channel speech enhancement has a long research history that has led to a variety of solutions like the classic single-channel Wiener filter \cite[Sec 11.4]{vary2006digital} or other estimators derived in a statistical framework \cite{1984ephraimSpeechEnhancementUsing, lotter2005SpeechEnhancementMAP, erkelens2007MinimumMeanSquareError}. Many recent advances in single-channel speech enhancement are driven by the modeling capabilities of \acp{DNN} \cite{xu2015aregression, rethage2018WavenetSpeechDenoising, park2017FullyConvolutionalNeural, pandey2019TCNNTemporalConvolutional}.

\begin{figure}
    \begin{minipage}[]{0.1\linewidth}
        \subcaption{}\label{fig:1-separated}
    \end{minipage}
    \begin{adjustbox}{minipage=0.85\linewidth}
        \centering
        \includegraphics{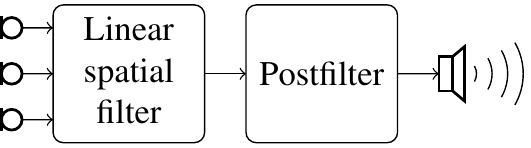}
    \end{adjustbox}
    
    \begin{minipage}[]{0.1\linewidth}
        \subcaption{}\label{fig:1-jointfilter}
    \end{minipage}
    \begin{adjustbox}{minipage=0.85\linewidth}
        \centering
        \includegraphics{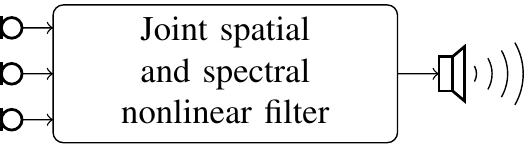}
    \end{adjustbox}
    \caption{(a) Illustration of the commonly employed two-step processing using a linear spatial filter (beamformer) followed by a single-channel postfilter. (b) Illustration of the nonlinear spatial filter investigated in this paper, which joins the spatial and spectral processing into a non-separable nonlinear operation.}
    \label{fig:1-comparison}
\end{figure}

It seems convenient to independently develop a spatial filter and a postfilter and combine them into a two-step procedure afterward as shown in Figure \ref{fig:1-separated}. If the noise follows a Gaussian distribution, this approach can even be regarded as optimal in the \ac{MMSE} sense as Balan and Rosca \cite{balan2002MicrophoneArraySpeech} have shown that the \ac{MMSE} solution can always be separated into the linear \ac{MVDR} beamformer and a postfilter. However, this separability into a linear spatial filter and a postfilter only holds under the restrictive assumption that the noise is Gaussian distributed. The work of Hendriks et al. \cite{hendriks2009OptimalMultichannelMeanSquared} points out that the \ac{MMSE} optimal solution for non-Gaussian noise joins the spatial and spectral processing into a single nonlinear operation. Throughout this work, we call such an approach a \emph{nonlinear spatial filter} for brevity even though spectral processing steps are also included. An illustration is given in Figure \ref{fig:1-jointfilter}.

The result of Hendriks et al. reveals that the common two-step multichannel processing scheme cannot be considered optimal for more general noise distributions than a Gaussian distribution. This leads to the question if we should invest in the development of nonlinear spatial filters for example using \acp{DNN}. Today, single-channel approaches often use the possibilities of \acp{DNN} to learn complex nonlinear estimators directly from data. In contrast, the field of multichannel speech enhancement is dominated by approaches that use \acp{DNN} only for parameter estimation of a beamformer \cite{heymann2016nnestimation, 2016xiaoDeepBeamforming} or restrict the network architecture in a way that a linear spatial processing model is preserved \cite{Sainath2017deepnns}. Only a few approaches with and without \acp{DNN} \cite{2019leeEndtoEndMultiChannelSpeech, 2020tolooshamsChannelAttentionDenseUNet, 2019itzhakNonlinearKroneckerProduct}  have been proposed that extend the spatial processing model to be nonlinear. Still, the questions of how much we can possibly gain by doing this, in which situations, and also where the benefit of using a nonlinear spatial filter comes from have not been addressed adequately. These are the questions that we aim to investigate in this paper.

This work is based on a previous conference publication \cite{2019teschNonlinearSpatialFiltering}. In \cite{tesch2020nonlinearspatial} we have studied related aspects of these questions. Here, we extend our previous work by more detailed derivations and new analyses that provide some insight into the functioning of the nonlinear spatial filter. In Section \ref{sec:3-theory}, we provide a detailed overview of the theoretical results from a statistical perspective. We include the previously outlined results and also provide a new simplified proof for the finding of Balan and Rosca in \cite{balan2002MicrophoneArraySpeech}. We then evaluate the performance benefit of a nonlinear spatial filter for heavy-tailed noise in Section \ref{sec:4-a-heavy-tailed}, for an inhomogeneous noise field created by five interfering human speakers in Section \ref{sec:4-b-inh-noise-speakers}, and real-world noise recordings in Section \ref{sec:4-d-chime}. In Section \ref{sec:5-interpretation}, we investigate the improved exploitation of spatial information by the nonlinear spatial filter and discuss practical issues of the used analytic nonlinear spatial filter.
Even though nonlinear spatial filters would most likely be implemented using \acp{DNN} in the future, in our analyses we rely on statistical \ac{MMSE} estimators to provide more general insights than by using \ac{DNN}-based nonlinear spatial filters which would be highly dependent on the network architecture and training data. 

\section{Assumptions and notation}
We assume that the signals recorded by a $\numMics$-dimensional microphone array decompose into a target speech and a noise component. For each microphone-channel $\micidx\in\{1,...,\numMics\}$, we segment the time-domain signal into overlapping windows and transform the signal to the frequency domain using the \ac{DFT} to obtain the \ac{DFT} coefficients $\randY_\micidx (\freqbinidx,\timeframeidx)$ with frequency-bin index $\freqbinidx$ and time-frame index \timeframeidx. Throughout this work, we use segments of length 32 ms with 16 ms shift and apply the square-root Hann function for spectral analysis and synthesis. By the additive signal model, the noisy \ac{DFT} coefficient can be written as the sum of the clean speech and the noise \ac{DFT} coefficients $\randS_\micidx(\freqbinidx, \timeframeidx)\in\mathbb{C}$ and $\randN_\micidx(\freqbinidx, \timeframeidx)\in\mathbb{C}$, i.e.,
\begin{IEEEeqnarray}{rCl}
\randY_{\micidx}(\freqbinidx, \timeframeidx) &=& \randS_{\micidx}(\freqbinidx, \timeframeidx) + \randN_{\micidx}(\freqbinidx, \timeframeidx).
\end{IEEEeqnarray}
As we model the \ac{DFT} coefficients to be random variables and assume independence with respect to the frequency-bin and time-frame index, we drop the indices $(\freqbinidx, \timeframeidx)$ to simplify the notation. We indicate random variables with uppercase letters and use lowercase letters for their respective realization. Furthermore, we assume all \ac{DFT} coefficients to be zero-mean and speech and noise to be uncorrelated.

We stack the noisy and noise \ac{DFT} coefficients into vectors $\brandY=[\randY_1, \randY_2, ..., \randY_\ell]^T\in\mathbb{C}^\numMics$ and $\brandN=[\randN_1, \randN_2, ..., \randN_\ell]^T\in\mathbb{C}^\numMics$and obtain the vector of speech \ac{DFT} coefficients $\brandS\in\mathbb{C}^\numMics$ by multiplying the clean speech signal coefficient $\randS\in \mathbb{C}$ with the so-called steering vector $\steeringvec\in\mathbb{C}^\numMics$, which accounts for the propagation path between the target speaker and the microphones. We can then rewrite the signal model as 
\begin{IEEEeqnarray}{rCl}
\brandY &=& \steeringvec\randS + \brandN.
\end{IEEEeqnarray}
The noise correlation matrix is denoted by $\corN = \mathbb{E}[\brandN\brandN^H]\in\mathbb{C}^{D\times D}$ with the statistical expectation operator $\mathbb{E}$ and $(\cdot)^H$ denoting the Hermitian transpose. The spectral power of the target speech signal is given by $\psdS=\mathbb{E}[|\randS|^2]\in\mathbb{R}^+$. When appropriate, we use the polar representation for complex-valued quantities, e.g., $\realS=|\realS|e^{j\varphi_\realS}\in\mathbb{C}$, and then let $\varphi$ denote the phase of the complex number.  

\section{Linearity of the optimal spatial filter}
\label{sec:3-theory}
In this section, we aim to provide a more complete picture of the nature of the optimal spatial filter by aggregating existing results and presenting more straightforward derivations for some of these.
We identify the noise distribution as the key to linearity versus non-linearity of the spatial filter and also to the separability of spatial and spectral processing. Accordingly, in our considerations we distinguish the two cases of Gaussian distributed noise and non-Gaussian distributed noise or, more precisely, noise that follows a Gaussian mixture distribution.
\subsection{Gaussian Noise}
We start with revisiting the results from Balan and Rosca \cite{balan2002MicrophoneArraySpeech} and then provide a simplified proof that may be easier to follow. 
We assume that the vector of noise \ac{DFT} coefficients $\brandN$ follows a multivariate complex Gaussian distribution with zero mean and covariance matrix $\corN$, i.e., $\brandN \sim \complexGaussian (0,\corN)$. As we employ an additive signal model, the conditional distribution of the noisy \ac{DFT} coefficient vector $\brandY$ given information on the reference clean speech \ac{DFT} is a multivariate complex Gaussian distribution centered around the vector of clean speech \ac{DFT} coefficients $\steeringvec\realS$ with the same covariance matrix $\corN$. The corresponding conditional \ac{PDF} is given by \cite[Thm. 15.1]{kay2009FundamentalsStatisticalSignal} 
\begin{IEEEeqnarray}{rCl}\label{eq:likelihood}
p_{\brandY}(\brealY | \realS) &=& \dfrac{1}{\pi^D |\corN|}\exp\left\{- (\brealY -\steeringvec \realS)^H \invCorN (\brealY - \steeringvec \realS)\right\}.
\end{IEEEeqnarray}
Our goal is to show that the linear \ac{MVDR} beamformer defined as 
\begin{IEEEeqnarray}{rCl}
\T{MVDR}(y) = \dfrac{\steeringvec^H\invCorN\brealY}{\steeringvec^H\invCorN\steeringvec}
\end{IEEEeqnarray}
is the optimal spatial filter with respect to the \ac{MAP}, \ac{MMSE} and \ac{ML} optimization criterion if the noise follows a Gaussian distribution.

Balan and Rosca \cite{balan2002MicrophoneArraySpeech} rely on the concept of sufficient statistics to prove the property in question for the \ac{MMSE} optimization criterion. In our context, the MVDR beamformer $\T{MVDR}$ is a \emph{sufficient statistic in the Bayesian sense} if 
\begin{IEEEeqnarray}{rCl}\label{eq:suffstat}
p_{\randS}(\realS|\brealY) & = & p_{\randS}(\realS| \T{MVDR}(\brealY))
\end{IEEEeqnarray}
holds for every observation $\brealY$ and every prior distribution of $\randS$ \cite[Thm. 2.4]{schervish1995TheoryStatistics}. We infer from (\ref{eq:suffstat}) that all information about $\randS$ contained in the noisy observation is retained in the output of the \ac{MVDR} beamformer despite the fact that the \ac{MVDR} beamformer reduces the dimension of the multidimensional input to one dimension. Note that the variable of interest $\randS$ in the above definition is a random variable. In contrast, $\T{MVDR}$ is a \emph{sufficient statistic in the classical sense} for the true clean speech \ac{DFT} coefficient $\realS$, which is not assumed to be a random variable, if the conditional distribution of the noisy observation $\brandY$ given $\T{MVDR}(\brandY)$ does not depend on $\realS$ \cite[Def. IV.C.1]{poor1994IntroductionSignalDetection}.

As a first step, Balan and Rosca deduce that the \ac{MVDR} beamformer is a sufficient statistic in the classical sense from the Fisher-Neyman factorization theorem \cite[Prop. IV.C.1]{poor1994IntroductionSignalDetection}\cite[Cor. 2.6.1]{lehmann2005TestingStatisticalHypotheses}, which is applicable since the conditional \ac{PDF} of the observation $\brandY$ given $\randS$ in (\ref{eq:likelihood}) can be rewritten as
\begin{IEEEeqnarray*}{rCl}\label{eq:likelihoodfactorized}
p_{\brandY}(\brealY |\realS) &=& \underbrace{\frac{1}{\pi^D |\corN|}\exp\{-\brealY^H\invCorN\brealY\}}_{h(\brealY )}\\
&&\times\underbrace{\exp\left\{\steeringvec^H\invCorN\steeringvec \left(2 \Real\left\{\realS^*\T{MVDR}(\brealY )\right\}- |\realS|^2 \right)\right\}}_{g(\realS, \T{MVDR}(\brealY ))}\\
&=& h(\brealY)g(\realS, \T{MVDR}(\brealY ))\\
&=& h(\brealY)g(\realS, \realZ).\IEEEyesnumber
\end{IEEEeqnarray*}
under the Gaussian noise assumption. In the last line of equation (\ref{eq:likelihoodfactorized}), we replaced the random variable $\T{MVDR}(\brandY)$ with $\randZ$, i.e., 
\begin{IEEEeqnarray}{rCl}\label{eq:substitude}
\randZ &=&\T{MVDR}(\brandY),\IEEEyesnumber
\end{IEEEeqnarray}
and will now continue to use this substitute when it improves the readability. In a second step, Balan and Rosca conclude that the \ac{MVDR} beamformer is a sufficient statistic in the Bayesian sense because any statistic that is sufficient in the classical sense is also sufficient in the Bayesian sense \cite[Thm. 2.14.2]{schervish1995TheoryStatistics}. 

We now provide a proof of the $\T{MVDR}$ being a sufficient statistic of $\randS$ in the Bayesian sense, which does not require a reference to advanced stochastic theorems. For this, we compute a factorization of the likelihood \ac{PDF} of $\randZ$ $p_\randZ(\realZ|\realS)$ with $\randZ$ defined in (\ref{eq:substitude}) as the output of the \ac{MVDR} beamformer for the noisy input $\brandY$. From the properties of the multivariate complex Gaussian distribution undergoing a linear transformation \cite[Appx. 15B]{kay2009FundamentalsStatisticalSignal}, we infer that $\randZ$ given $\randS$ is distributed according to a one-dimensional complex Gaussian distribution with mean $\realS$ and variance $(\steeringvec^H\invCorN\steeringvec)^{-1}$, i.e., 
\begin{IEEEeqnarray}{rCl}\label{eq:ltofgaussian}
p_\randZ(\realZ|\realS) &=& \complexGaussian \left(\realS, (\steeringvec^H\invCorN\steeringvec)^{-1}\right).
\end{IEEEeqnarray}
The corresponding \ac{PDF} at the output of the beamformer can be factorized as
\begin{IEEEeqnarray*}{rCl}\label{eq:outputlikelihoodfactorized}
p_\randZ(\realZ |\realS) &=& \underbrace{\frac{\steeringvec^H\invCorN\steeringvec}{\pi}\exp\{-\brealY^H\invCorN\brealY\left|\realZ\right|\}}_{f(\brealY )}\\
\IEEEeqnarraymulticol{3}{r}{
\times\underbrace{\exp\left\{\steeringvec^H\invCorN\steeringvec \left(2 \Real\left\{\realS^*\realZ\right\}- |\realS|^2 \right)\right\}}_{g(\realS, \realZ)}}\\
&=& f(\brealY)g(\realS, \realZ).\IEEEyesnumber
\end{IEEEeqnarray*}
Using (\ref{eq:likelihoodfactorized}) we rewrite the posterior distribution 
as
\begin{equation}
\begin{IEEEeqnarraybox}[][c]{rCl}
p_{\randS}(\realS|\brealY) &=& \dfrac{p(\brealY|\realS)p(\realS)}{\int_{\mathbb{C}}p(\brealY|\realS)p(\realS)ds}\\
&=& \dfrac{h(\brealY)g(\realS, \realZ)p(\realS)}{\int_{\mathbb{C}}h(\brealY)g(\realS, \realZ)p(\realS)ds}.
\end{IEEEeqnarraybox}
\end{equation}
Since the term $h(\brealY)$ in the denominator does not depend on the integration variable $\realS$, this term cancels with the corresponding term in the numerator. Next, we extend the fraction with the term $f(\brealY)$ from (\ref{eq:outputlikelihoodfactorized}) to obtain
\begin{equation} 
\begin{IEEEeqnarraybox}[][c]{rCl}
p_\realS(\realS|\brealY) &=&  \dfrac{f(\brealY)g(\realS, \realZ)p(\realS)}{\int_{\mathbb{C}}f(\brealY)g(\realS, \realZ)p(\realS)ds}\\
&=& p_\randS(\realS |\realZ)\\
&=& p_\randS(\realS |\T{MVDR}(\brealY)),
\end{IEEEeqnarraybox}
\end{equation}
which is the identity we wanted to prove (cf. (\ref{eq:suffstat})). 
Consequently, as the posterior given the noisy observation $\brealY$ equals the posterior given the output of the MVDR beamformer $\T{MVDR}(\brealY)$, we find that 
\begin{IEEEeqnarray}{rCl}\label{eq:map}
\T{MAP}(\brealY) 
&=& \argmax_{\realS\in\mathbb{C}} p_\randS(\realS|\T{MVDR}(\brealY))
\end{IEEEeqnarray}
holds. The \ac{MVDR} beamformer reduces its multidimensional input to a single-channel output and, therefore, the right-hand side of (\ref{eq:map}) can be seen as a single-channel postfilter working on the output of the \ac{MVDR} beamformer. Since the \ac{MMSE} estimator complies with the mean of the posterior, a similar decomposition in a linear spatial filter and a spectral postfilter is given by
\begin{equation}
\begin{IEEEeqnarraybox}[][c]{rCl}
\T{MMSE}(\brealY) &=& \mathbb{E}[\randS|\brealY]\\
&=& \mathbb{E}[\randS|\T{MVDR}(\brealY)].
\end{IEEEeqnarraybox}
\end{equation}

Because the relationship (\ref{eq:suffstat}) holds for all prior distributions of $\randS$, a decomposition of the \ac{MAP} and \ac{MMSE} estimators into a linear spatial filter followed by a postfilter exist independently from any further assumptions regarding the prior distribution of the clean speech \ac{DFT} coefficient. 

Finally, we consider the \ac{ML} estimator. Starting from (\ref{eq:likelihoodfactorized}) and exploiting the monotony of the logarithm and Euler's formula, we find the representation 
\begin{IEEEeqnarray*}{rCl}
\T{ML}(\brealY) &=& \argmax_{\realS\in\mathbb{C}} p_{\brandY} (\brealY|\realS)\\
&=& \argmax_{\realS\in\mathbb{C}} 2 \Real \{\realS^*\underbrace{\T{MVDR}(\brealY )}_{=z}\}- |\realS|^2\IEEEyesnumber\\
&=& \argmax_{\realS\in\mathbb{C}} 2\cdot |\realS|\cdot |z|\cdot \cos(\varphi_{z} - \varphi_\realS) - |\realS|^2.
\end{IEEEeqnarray*}
Clearly, this function is maximized when the phase of $\realS$ matches the phase of $\T{MVDR}(\brealY)$, as then the cosine function is maximized. Equating the derivative with respect to $|s|$ to zero and solving for $|s|$ reveals that the magnitude of the \ac{MVDR} beamformer maximizes the likelihood. Thus, $\T{ML}(\brealY) = \T{MVDR}(\brealY)$, i.e. the MVDR beamformer is the maximum likelihood estimator  of the clean speech \ac{DFT} coefficient as also stated in \cite[Sec. 6.2.1.2]{trees2004OptimumArrayProcessing}.

\subsection{Non-Gaussian Noise}
As we have seen, if the noise \ac{DFT} coefficients follow a Gaussian distribution, then a linear spatial filter can be considered optimal. However, Hendriks et al. \cite{hendriks2009OptimalMultichannelMeanSquared} have shown that this does not need to be the case for non-Gaussian distributed noise. In their work, they model the noise distribution with a multivariate complex Gaussian mixture distribution. The $M$ Gaussian mixture components with respective covariance matrix $\corM$, $m\in\{1,...,M\}$, are assumed to be zero-mean such that the conditional PDF given the clean speech is given by \begin{IEEEeqnarray*}{rCl}\label{eq:mixturelikelihood}
p_{\brandY}(\brealY | \realS) &=& \sum_{m=1}^M c_m \complexGaussian (\steeringvec\realS, \corM).\IEEEyesnumber
\end{IEEEeqnarray*}
with mixture weights $c_m$ that sum to one. 
Hendriks et al. assume that the amplitude $A_\randS$ and phase $\varphi_\randS$ of the clean speech \ac{DFT} coefficient are independent. They model the phase to be uniformly distributed over the interval $[0,2\pi)$ and assume the amplitude to be generalized-Gamma distributed (\cite[Eq. 1]{erkelens2007MinimumMeanSquareError}, with $\gamma=2$ and $\beta=\nu/\psdS$). The corresponding \ac{PDF} 
\begin{IEEEeqnarray*}{rCl}\label{eq:generalizedGammaAmplitude}
p_{A_\randS}(\realA) &=& 2\dfrac{\left( \dfrac{\nu}{\psdS}\right)^\nu}{\Gamma(\nu)} a^{2\nu-1}\exp\left\{-\frac{\nu}{\psdS}a^2\right\} \text{ with } \nu> 0,~a \geq 0\\\IEEEyesnumber
\end{IEEEeqnarray*}
depends on the speech shape parameter $\nu$, and $\Gamma (\cdot)$ is the Gamma function.
Under these assumptions, Hendriks et al. derive the MMSE estimator
\begin{IEEEeqnarray*}{rCl}
\T{MMSE}(\brealY )&=& \nu\frac{{\displaystyle\sum_{m=1}^M} \frac{c_mQ_m}{|\corM |}e^{\left[-\brealY^H\invCorM\brealY\right]}\frac{\psdS T_{\text{MVDR}}^{(m)}(\brealY)\confhypergeom (\nu+1,2,P_m)}{\nu(\steeringvec^H\invCorM\steeringvec)^{-1}+\psdS}}
{{\displaystyle \sum_{m=1}^M} \frac{c_mQ_m}{|\corM|}e^{\left[-\brealY^H\invCorM\brealY\right]}\confhypergeom (\nu,1,P_m)}\\*\IEEEyesnumber\label{eq:mmsegmm}\\
\noalign{\noindent with }
\IEEEeqnarraymulticol{3}{C}{
T_{\text{MVDR}}^{(m)}(\brealY) = \dfrac{\steeringvec^H\invCorM\brealY}{\steeringvec^H\invCorM\steeringvec},\quad
Q_m = (\nu + \steeringvec^H\invCorM\steeringvec\psdS )^{-\nu},}\\
\IEEEeqnarraymulticol{3}{C}{
\text{and}\quad P_m = \dfrac{\psdS\steeringvec^H\invCorM\steeringvec \left|T_{\text{MVDR}}^{(m)}(\brealY)\right|^2}{\nu (\steeringvec^H\invCorM\steeringvec)^{-1}+\psdS}}
\end{IEEEeqnarray*}
with $\confhypergeom(\cdot,\cdot,\cdot)$ being the confluent hypergeometric function \cite[Sec. 9.21]{gradshteyn2000TableIntegralsSeries}. From (\ref{eq:mmsegmm}) it is apparent that the \ac{MMSE} estimator \emph{cannot} be decomposed in a linear spatial filter and a spectral postfilter. This is because the linear term $T_{\text{MVDR}}^{(m)}$ as well as the quadratic term $\brealY
^H \invCorM\brealY$ depend on the summation index $m$. The spatial nonlinearity is particularly evident from the aforementioned quadratic term.

Throughout this work, we compare the results of the optimal spatially nonlinear MMSE estimator with a classical setup comprised of a linear spatial filter and (nonlinear) spectral postfilter. Figure \ref{fig:1-comparison} provides an illustration of the compared estimators: part (b) represents the nonlinear spatial filter $\T{MMSE}$ given in (\ref{eq:mmsegmm}) and part (a) corresponds to a combination of the \ac{MVDR} beamformer with an \ac{MMSE}-optimal postfilter. We now derive the postfilter under the same distributional assumptions as $\T{MMSE}$. 

Since the \ac{MVDR} beamformer is linear, we can infer the distribution of the beamformer output and observe that it follows a one-dimensional complex Gaussian mixture distribution with \ac{PDF}
\begin{IEEEeqnarray}{rCl}
p(\T{MVDR}(\brealY) | \realS) &=& \sum_{m=1}^M c_m \mathcal{N}_\mathbb{C}\bigg(\realS, \underbrace{\dfrac{\steeringvec^H\invCorN\corM\invCorN\steeringvec}{(\steeringvec^H\invCorN\steeringvec)^2}}_{\varM})\bigg)\IEEEeqnarraynumspace
\end{IEEEeqnarray}
for an input $\brandY$ that is distributed according to a multivariate complex Gaussian mixture distribution. The Gaussian mixture components have the mean $\realS$ and variance $\varM$, $m\in\{1,...,M\}$.  Based on this observation, we compute the \ac{MMSE}-optimal spectral postfilter using \cite[Eq. 3.339, Eq. 6.643.2, Eq. 9.220.2]{gradshteyn2000TableIntegralsSeries} and \cite[Eq. 10.32.3]{nist2018} and obtain the estimator 
\begin{IEEEeqnarray*}{rCl}
\T{MVDR-MMSE}(\brealY )&=\\
\IEEEeqnarraymulticol{3}{C}{
\nu\frac{{\displaystyle\sum_{m=1}^M} \frac{c_mQ_m}{\varM}e^{\left[-\frac{|\T{MVDR}(\brealY)|^2}{\varM}\right]}\frac{\psdS\T{MVDR}(\brealY)\confhypergeom (\nu+1,2,P_m)}{\nu\varM+\psdS}}
{{\displaystyle\sum_{m=1}^M} \frac{c_mQ_m}{\varM}e^{\left[-\frac{|\T{MVDR}(\brealY)|^2}{\varM}\right]}\confhypergeom (\nu,1,P_m)}}\IEEEeqnarraynumspace\IEEEyesnumber\label{eq:mmsemvdrgmm}\\
\noalign{\noindent with }
\IEEEeqnarraymulticol{3}{C}{
\corN= \sum_{m=1}^M c_m \corM,\quad
\varM= \dfrac{\steeringvec^H\invCorN\corM\invCorN\steeringvec}{(\steeringvec^H\invCorN\steeringvec)^2},}\\
\IEEEeqnarraymulticol{3}{C}{
Q_m = (\dfrac{1}{\varM}+\dfrac{\nu}{\psdS})^{-\nu}
\quad\text{and}\quad
P_m = \dfrac{\psdS\invVarM |\T{MVDR}(\brealY)|^2}{\nu \varM + \psdS}.}
\end{IEEEeqnarray*}
that sequentially combines linear spatial processing with \ac{MMSE}-optimal spectral postprocessing as depicted in Figure \ref{fig:1-separated}.

\section{Evaluation of the benefit of a nonlinear spatial filter in non-Gaussian noise}
\label{sec:4-sampled-data}
Section \ref{sec:3-theory} points out that using a nonlinear spatial filter is \ac{MMSE}-optimal and, thus, may be beneficial if the noise does not follow a Gaussian distribution. It is well known that the \ac{DFT} coefficients of speech are often better modeled by a more heavy-tailed distribution than a Gaussian if originating from \ac{STFT} segments with short duration \cite{lotter2005SpeechEnhancementMAP,martin2005SpeechEnhancementBased}. Consequently, one may argue that this as well applies to noise \ac{DFT} coefficients if the background noise stems from human speakers. In any case, Martin \cite{martin2005SpeechEnhancementBased} observed that heavy-tailed distributions also provide a good fit for \ac{DFT} coefficients of some types of noise in the one-dimensional case.

In this section, we investigate the potential of the optimal nonlinear spatial filter versus the classical separated setup with a linear spatial filter and a spectral postfilter for noise with a non-Gaussian distribution. Section \ref{sec:4-a-heavy-tailed} presents our findings for noise that departs from Gaussianity by means of heavier tails but with a rather simple spatial structure. We published parts of this analysis and of the analysis in Section~\ref{sec:4-d-chime} in \cite{2019teschNonlinearSpatialFiltering}. However, here we also include the multichannel Wiener filter for comparison, compute more detailed performance metrics, and have made changes to the speech power parameter estimation scheme. In Section \ref{sec:4-b-inh-noise-speakers} we provide results for noise that is modeling a spatially more diverse noise field created by five interfering human speakers and in Section \ref{sec:4-d-chime} we evaluate the nonlinear spatial filtering approach based on real-world noise recordings from the CHiME3 database. Please find audio examples for all experiments on our website\footnote{\label{link:audioexamples}\url{https://uhh.de/inf-sp-nonlinear-spatial-filter-tasl2021}}.

\subsection{Heavy-tailed noise distribution}\label{sec:4-a-heavy-tailed}
In our first experiment, we investigate the performance of the nonlinear spatial filter $\T{MMSE}$ by mixing the target speech signal at the microphones with multichannel noise that is sampled from a heavy-tailed Gaussian mixture distribution.
\subsubsection{Noise distribution model}
We construct a Gaussian mixture distribution with an adjustable heavy-tailedness by combining Gaussian components with scaled versions of the same covariance matrix. Therefore, we set the $m$th mixture component's covariance matrix $\corM$ to be 
\begin{IEEEeqnarray}{rCltrCl}
\corM &=& \dfrac{b^{m-1}}{r} \corN &\quad with \quad & r&=&\sum_{m=1}^M c_m b^{m-1} 
\label{eq:scaledGM}
\end{IEEEeqnarray}
and scaling factor $b\in\mathbb{R}^+$. The constant $r$ ensures correct normalization such that the overall covariance matrix of our scaled Gaussian mixture distribution remains $\corN$.

We rely on the kurtosis to quantify the heavy-tailedness of the scaled Gaussian mixture distributions. It is a statistical measure that accounts for the likelihood of the occurrence of outliers \cite{westfall2014KurtosisPeakedness1905} and it has been extended for real-valued multivariate distributions by Mardia \cite{mardia1970MeasuresMultivariateSkewness}. We extend it to complex-valued random vectors $\brandX\in\mathbb{C}^n$ with mean $\vec{\mu}$ and covariance $\covX$ by defining its kurtosis to be 
\begin{IEEEeqnarray}{rCl}\label{eq:complexkurtosis}
\kappa_\mathbb{C}(\brandX) &=& \mathbb{E}\left[(2(\brandX - \vec{\mu})^H\invCovX(\brandX - \vec{\mu}))^2\right].
\end{IEEEeqnarray}
A complex-valued $n$-dimensional Gaussian distribution can equivalently be formulated as a real-valued $2n$-dimensional Gaussian distribution \cite[Thm. 15.1]{kay2009FundamentalsStatisticalSignal}. The additional factor of two in (\ref{eq:complexkurtosis}) ensures that the same kurtosis value results for both formulations of the same distribution.
Using \cite[Sec. 8.2.4]{petersen2012matrixcookbook}, we compute the kurtosis of a vector $\brandN$ distributed according to a scaled Gaussian mixture distribution to obtain
\begin{IEEEeqnarray}{rCl}
\kappa_\mathbb{C}(\brandN) = 2D(2+2D)\underbrace{\sum_{m=1}^M c_m \dfrac{b^{2(m-1)}}{r^2}}_{q}\label{eq:scaledGaussiankurtosis}
\end{IEEEeqnarray}
and observe that $\kappa_\mathbb{C}(\brandN)$ is given by the kurtosis of a $D$-dimensional complex Gaussian distribution multiplied by a factor $q$ that depends on the scaling factor $b$ and the number of mixture components $M$.

\subsubsection{Experimental setup}
In our test scenario, we use five microphones arranged in a linear array with 5 cm spacing and broadside orientation towards the target signal source and model the propagation path between the target speaker and the microphones based on time delays only. We perform the evaluation using 48 clean speech signals taken from the WSJ0 dataset \cite{wsjdata2007} that are balanced between female and male speakers. The noise DFT coefficients are samples from a scaled Gaussian mixture distribution with scale factor $b=2$ and a variable number of mixture components with equal weight $c_m=\frac{1}{M}$, $m\in\{1,...,M\}$. The noise covariance matrix $\corN$ models a diffuse noise field with a small portion (factor of 0.05) of additional spatially and spectrally white noise as in \cite[Eq. 27]{pan2014perfstudyofMVDRsourceangle}.
The noise and speech are combined such that a \ac{SNR} of 0 dB is obtained. 

\subsubsection{Performance evaluation} 
\begin{figure}
    \centering
    \includegraphics{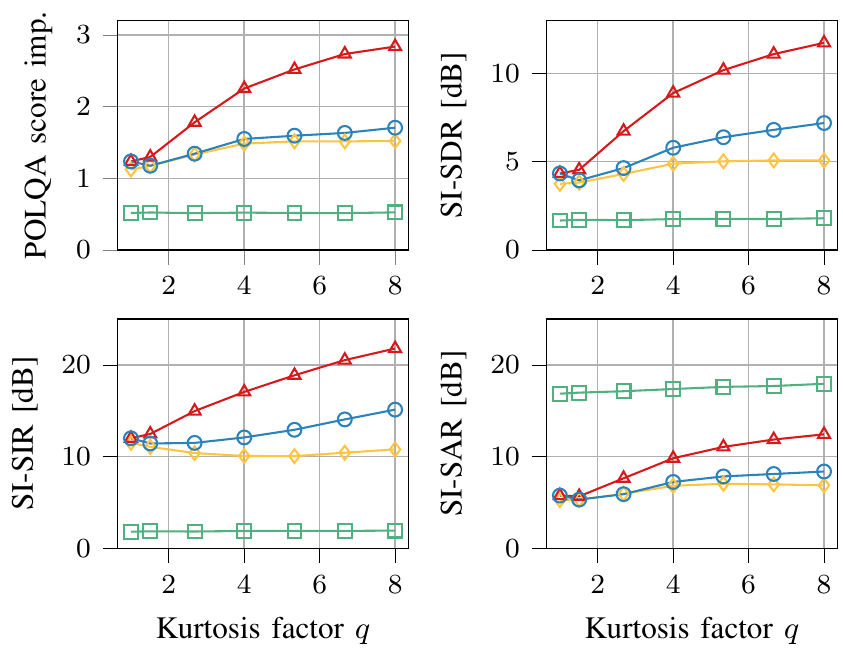}
    \includegraphics{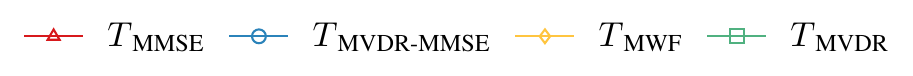}
    \caption{\ac{POLQA}, \ac{SI-SDR}, \ac{SI-SIR} and \ac{SI-SAR} results for scaled Gaussian mixture noise distributions with increased heavy-tailedness in diffuse noise. }\label{fig:4-a-heavy-tailed}
\end{figure}

Figure \ref{fig:4-a-heavy-tailed} provides a performance comparison of the jointly spatial and spectral nonlinear $\T{MMSE}$ and the spatially linear $\T{MVDR-MMSE}$ with a nonlinear postfilter. The speech shape parameter is set to $\nu = 0.25$ for both estimators. Furthermore, we display results obtained with the well-known linear spatial filter $\T{MVDR}$ without a postfilter and the multichannel Wiener filter $\T{MWF}$, which is the \ac{MMSE}-optimal solution if noise \emph{and} speech follow a Gaussian distribution, i.e., $\T{MVDR-MMSE}$ with $\nu=1$ and $M=1$. The performance results are displayed with respect to the kurtosis factor $q$ on the x-axis indicating an increased heavy-tailedness of the noise distribution from left to right.

The plot in the upper left corner shows the performance with respect to the improvement of the \ac{POLQA} measure \cite{polqa2018}, which is the successor of \ac{PESQ} \cite{pesq2007} and returns the expected \ac{MOS}, which takes values from one (bad) to five (excellent). In any plot of Figure \ref{fig:4-a-heavy-tailed}, we are particularly interested in the performance difference of the $\T{MMSE}$ (red) and $\T{MVDR-MMSE}$ (blue) as this gap characterizes the potential performance gain of a nonlinear spatial filter. For \ac{POLQA}, we observe an increase of the performance difference up to 1.1 \ac{POLQA} score improvement as the noise distribution shifts towards a more heavy-tailed distribution.

The estimators including a postfilter, $\T{MMSE}$, $\T{MVDR-MMSE}$, and $\T{MWF}$, require an estimate of the speech \ac{PSD} $\psdS$. In contrast to our previous paper \cite{2019teschNonlinearSpatialFiltering}, here we do not rely on oracle knowledge of the clean speech signal to estimate this parameter but obtain an estimate from the noisy signal based on the cepstral smoothing technique \cite{breithaupt2008NovelPrioriSNR}. This results in an increased performance gap between $\T{MMSE}$ and $\T{MVDR-MMSE}$. From this finding, we conclude that a nonlinear spatial filtering approach is even more beneficial if the performance of the postfilter decreases due to estimation errors of the spectral power of the target speech signal. 

The next three plots (upper right and second row) display the performance results for the \ac{SI-SDR}, \ac{SI-SIR}, and \ac{SI-SAR} measures as defined in \cite{roux2019SDRHalfbakedWell}. We compute the \ac{SI-SDR}, \ac{SI-SIR}, and \ac{SI-SAR} for segments of length 10 ms without overlap and include only segments with target speech activity similar to the computation of the segmental \ac{SNR} in \cite{2012gerkmannUnbiasedMMSEBasedNoise}. 
The performance results based on the \ac{SI-SDR} measure show a similar structure to the ones obtained with \ac{POLQA}. For high kurtosis values, we observe a performance gap of 4.5 dB for $\T{MMSE}$ and $\T{MVDR-MMSE}$. Furthermore, the difference between $\T{MVDR-MMSE}$ and $\T{MWF}$ for high kurtosis values, which results exclusively from the different postfilter, is more obvious. The observed performance gaps, in particular the performance advantage of the nonlinear spatial filter, coincide with our own listening experience\footref{link:audioexamples}.

For the computation of the \ac{SI-SIR} and \ac{SI-SAR} measure displayed in the second row, the residual noise is split into interference noise and artifacts. It is striking to see that the red graph of $\T{MMSE}$ runs above the blue graph of $\T{MVDR-MMSE}$ in both plots meaning that the nonlinear spatial filter achieves better noise reduction and fewer speech distortions at the same time. The better performance with respect to the \ac{SI-SAR} measure is quite notable as we can see that the linear \ac{MVDR} beamformer introduces very few speech distortions but its combination with different postfilters ($\T{MVDR-MMSE}$ and $\T{MWF}$) still performs worse than the joint spatial and spectral non-linear processing by $\T{MMSE}$.
\subsection{Inhomogeneous noise field (interfering speech)}\label{sec:4-b-inh-noise-speakers}
Instead of sampling a Gaussian mixture distribution as in the previous section or in \cite{tesch2020nonlinearspatial}, we now use a setup with five interfering point sources arranged as illustrated in Figure \ref{fig:expsetup} and, this way, move closer towards realistic noise scenarios. The estimators $\T{MMSE}$ and $\T{MVDR-MMSE}$ have been derived under a Gaussian mixture noise assumption. To be consistent with this modeling assumption, we require the five interfering point sources to not be Gaussian distributed or not be simultaneously active per time-frequency bin, because otherwise the overall noise resulting from the different interfering sources would also be Gaussian distributed. Choosing human speakers as interfering sources, this assumption is commonly assumed to hold and referred to as w-disjoint orthogonality \cite{yilmaz2004BlindSeparationSpeech}.

\subsubsection{Experimental setup}\label{subsec:4-b-expsetup}
\begin{figure}
    \centering
    \includegraphics{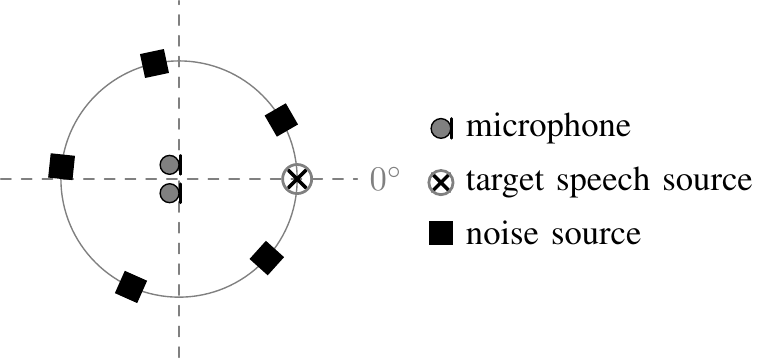}
    \caption{Illustration of the experiment setup with a two-dimensional linear microphone array, a target speech source in broadside direction and five interfering point sources (human speakers (Section \ref{sec:4-b-inh-noise-speakers}) or Gaussian bursts (Section \ref{sec:5-interpretation})).}
    \label{fig:expsetup}
\end{figure}

As can be seen in Figure \ref{fig:expsetup}, the target speech source is placed in the broadside direction of the two-dimensional linear microphone array with 6 cm microphone spacing. We sample the target speech signal and the interfering signals from distinct subsets of the WSJ0 dataset. The two-dimensional noise signal is then obtained by multiplying the interfering speech signals with the steering vectors $\steeringvec_i$, $i\in\{0,...,4\}$, and adding the individual interfering sources' signals. The steering vector $\steeringvec_i$ of the $i$th interfering source positioned at $\theta_i=\frac{\pi}{6}+\frac{2\pi}{5}i$ radians models the relative time difference of arrival at the microphones. The target speech signal and noise signal are rescaled to correspond to an \ac{SNR} of 0 dB.

\newcolumntype{m}[1]{D{:}{\pm}{#1}}
\begin{table}
\centering
\begin{tabular}{lm{-1}m{-1}}\toprule
 & \multicolumn{1}{c}{Interfering speech} & \multicolumn{1}{c}{Gaussian bursts}\\\midrule
$\Delta$ POLQA & 0.84~:~0.04 & 2.64~:~0.08 \\\arrayrulecolor{black!30}\midrule
$\Delta$ SI-SDR & 4.63~:~0.15 & 9.92~:~0.30 \\
$\Delta$ SI-SAR & 3.91~:~0.16 & 8.39~:~0.26 \\
$\Delta$ SI-SIR & 6.44~:~0.22 & 14.95~:~0.46 \\\arrayrulecolor{black!30}\midrule
ESTOI (noisy)& 0.49~:~0.01 & 0.57~:~0.02 \\
ESTOI ($\T{MMSE}$)& 0.85~:~0.01 & 0.94~:~0.00 \\
ESTOI ($\T{MVDR-MMSE}$)& 0.72~:~0.01 & 0.67~:~0.02 \\
\arrayrulecolor{black}\bottomrule
\end{tabular}
\caption{Performance results (mean and the 95\% confidence interval) of the $\T{MMSE}$ and $\T{MVDR-MMSE}$ estimators for an inhomogeneous noise field with interfering speech and Gaussian sources as described in Section \ref{sec:4-b-inh-noise-speakers} and Section \ref{sec:5-interpretation} respectively.}
\label{table:multisource-results}
\end{table}

We now compare the performance of the $\T{MMSE}$ and $\T{MVDR-MMSE}$ estimators for the inhomogeneous noise field. For this, we require estimates of the Gaussian mixture components' covariance matrices $\corM$ and the mixture weights $c_m$. We choose the number of components equal to the number of interfering sources, i.e., $M=5$, and estimate the Gaussian mixture parameters using the \ac{EM} algorithm \cite{bishop2006pattern} applied to overlapping signal segments of length 250 ms and with an overlap of 50\% from the pure noise signal. As before, we estimate the spectral power of speech using the cepstral smoothing technique and use a speech shape parameter $\nu=0.25$.

\subsubsection{Performance evaluation}
The first column of Table \ref{table:multisource-results} displays the performance results for the described simulation. For the performance measures in the first four rows, preceded with a $\Delta$ symbol, we report the performance difference between $\T{MVDR-MMSE}$ and $\T{MMSE}$ averaged over 48 samples. We observe that the nonlinear spatial filter delivers a considerable performance gain that amounts to 4.63 dB \ac{SI-SDR} and a POLQA score of 0.84. The bottom part of Table \ref{table:multisource-results} presents ESTOI \cite{2016jensen} scores for the noisy signal and the enhancement results obtained with $\T{MMSE}$ and $\T{MVDR-MMSE}$. The ESTOI scores provide a measure of speech intelligibility. As for the other performance measures, we find that the nonlinear spatial filter outperforms the combination of a linear spatial filter and a postfilter as the $\T{MMSE}$ estimator yields an ESTOI score of 0.85 as opposed to the result of 0.72 achieved by $\T{MVDR-MMSE}$. 
\subsection{Real-world CHiME3 noise}\label{sec:4-d-chime}
Furthermore, we investigate the performance of the nonlinear spatial filter for real-world noise from the CHiME3 database \cite{Barker2015CHiME} that has been recorded in four different environments: a cafeteria, a moving bus, next to a street, and in a pedestrian area.

\subsubsection{Experimental setup}
The CHiME3 data has been recorded using six microphones that are attached to a tablet computer. For this experiment, we use the simulated training subset of the official dataset, which has been created by mixing the recording of real-world background noise with a spatialized version of WSJ0 utterances. A detailed description of the data generation process can be found in \cite{Barker2015CHiME}. We evaluate on 48 randomly selected samples that are balanced between male and female speakers.

As before, we require an estimate of the time-varying Gaussian mixture distribution parameters and estimate them using oracle knowledge of the noise signal. For this, we apply the \ac{EM} algorithm to overlapping signal segments of length 750 ms. For both, $\T{MMSE}$ and $\T{MVDR-MMSE}$, we use $\nu=0.25$ and estimate the speech power $\psdS$ using the cepstral smoothing technique. In addition, we need to estimate the steering vector for the target speaker. For this, we employ oracle knowledge of the clean speech signal and extract the steering vector estimates as principal eigenvectors of the time-varying covariance matrix estimates obtained by recursive smoothing. 

\subsubsection{Performance evaluation}
\begin{figure}
    \centering
    \includegraphics{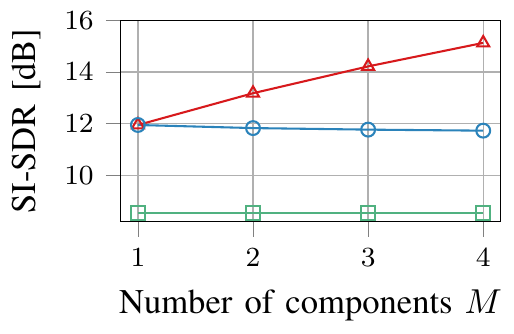}
    \includegraphics{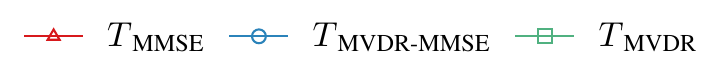}
    \caption{\ac{SI-SDR} results for cafeteria noise with respect to the number of mixture components used to fit the noise distribution.}\label{fig:4-c-chime}
\end{figure}

Again, we assess the performance gap between the nonlinear spatial filter $\T{MMSE}$ and the separated setup with a linear spatial filter and a postfilter $\T{MVDR-MMSE}$. Figure \ref{fig:4-c-chime} displays the \ac{SI-SDR} results for these estimators and also the \ac{MVDR} beamformer $\T{MVDR}$ with respect to the number of mixture components that have been fitted to a cafeteria background noise using the \ac{EM} algorithm. While the performance of the $\T{MMSE}$ estimator (red) improves with increased modeling capabilities of the mixture distribution, neither the $\T{MVDR}$ nor the $\T{MVDR-MMSE}$ estimator benefits from using more mixture components. As a result, we observe a performance gap of $3.17~\text{dB}$ \ac{SI-SDR} between the best result obtained with the nonlinear spatial filter and the best result obtained with the separated setup. This value coincides with the first entry of Table \ref{table:chime-results}. For the POLQA measure displayed in the second row, we find a difference of 0.59 POLQA score. The table shows bigger performance differences for the cafeteria (CAF) and pedestrian area (PED) noise than the bus (BUS) and street (STR) noise. We suppose that this reflects that the cafeteria and pedestrian area noise is less stationary as we hear the most significant differences for impulse like background noise. The ESTOI scores displayed at the bottom of Table \ref{table:chime-results} indicate that the nonlinear spatial filter is not only beneficial to the speech quality but also the speech intelligibility. Overall, we conclude that the Gaussian noise assumption does not seem to be valid for the examined real-world noise as the nonlinear spatial filter provides a notable benefit also for these recordings.

\begin{table}
\centering
\begin{tabular}{l@{}m{-1}@{}m{-1}@{}m{-1}@{}m{-1}}\toprule
& \multicolumn{1}{c}{CAF} & \multicolumn{1}{c}{BUS} & \multicolumn{1}{c}{PED} & \multicolumn{1}{c}{STR}\\ \midrule
$\Delta$ SI-SDR     &  3.17:0.19 &  2.48 : 0.26 & 3.31 : 0.24 &  2.07 : 0.28 \\\arrayrulecolor{black!30}\midrule
$\Delta$ POLQA		& 0.59 : 0.06 & 0.38 : 0.07 & 0.56 : 0.05  & 0.28 : 0.04 \\\arrayrulecolor{black!30}\midrule
ESTOI (noisy)    &  0.60:0.03 &  0.71 : 0.02 & 0.56 : 0.03 &  0.69 : 0.03 \\
ESTOI ($\T{MMSE}$)    &  0.94:0.01 &  0.97 : 0.01 & 0.93 : 0.01 &  0.96 : 0.01 \\
\parbox{2cm}{ESTOI ($\T{MVDR-MMSE}$)}  &  0.89:0.02 &  0.95: 0.01 & 0.88 : 0.02 &  0.94 : 0.01 \\\arrayrulecolor{black}\bottomrule 
\end{tabular}
\caption{Performance results (mean and the 95\% confidence interval) of the nonlinear spatial $\T{MMSE}$ and linear spatial filter combined with a postfilter $\T{MVDR-MMSE}$ for noise from the CHiME3 databse, which has been fitted with a Gaussian mixture distribution with four mixture components as described in Section \ref{sec:4-d-chime}.}
\label{table:chime-results}
\end{table}

\section{Interpretation: a nonlinear spatial filter enables superior spatial selectivity}
\label{sec:5-interpretation}
\begin{figure*}
    \begin{center}
        \includegraphics{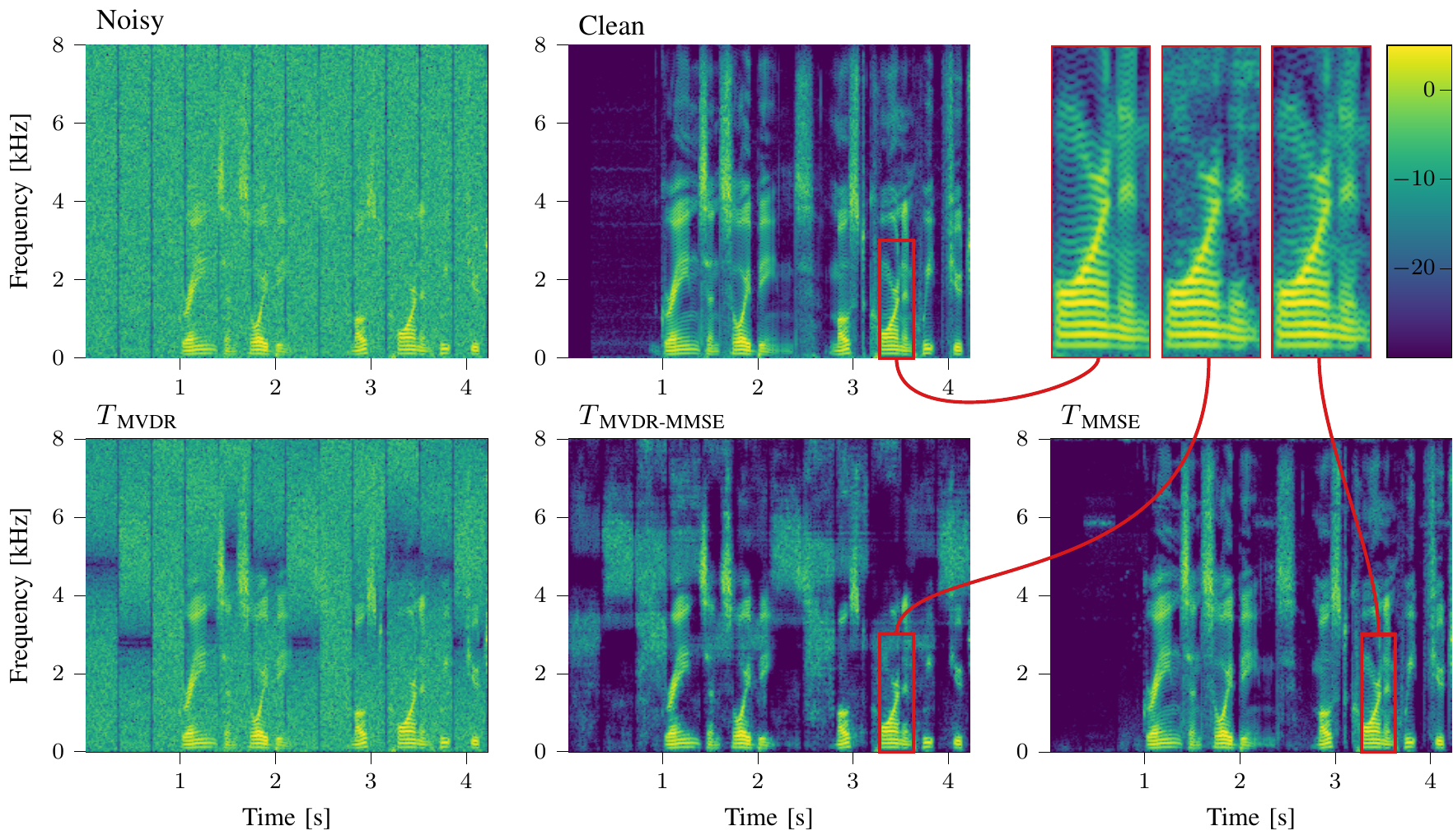}
    \end{center}
    \caption{Spectrograms of an example in an inhomogeneous noise field with five interfering sources emitting Gaussian noise bursts. The second row visualizes the processing results obtained with $\T{MVDR}$, $\T{MVDR-MMSE}$ and $\T{MMSE}$ and the top row shows the clean and noisy spectrograms as well as close-ups of the fine-structure of a voiced speech segment. }\label{fig:4-b-specgrograms}
\end{figure*}

We assume that the performance benefit of the nonlinear spatial filter reported in Section \ref{sec:4-b-inh-noise-speakers} and \ref{sec:4-d-chime} is due to the more efficient use of spatial information by the $\T{MMSE}$ estimator. Here we support this conjecture by an experiment that provides an insight into the functioning of the nonlinear spatial filter.
\subsubsection{Experimental setup}
We use the same geometric setup as described previously in Section \ref{sec:4-b-inh-noise-speakers} (Figure \ref{fig:expsetup}) but replace the interfering speech sources with sources that emit spectrally white Gaussian signals. To match the long-term non-Gaussianity assumption, only one interfering source emits a signal at a specific time instance. We implement this by using short (336 ms) non-overlapping Gaussian bursts for the interfering sources. The so created noise signal can be viewed as stationary regarding its spectral characteristics except at the segment boundaries. By applying the \ac{EM} algorithm to the full-length noise signal we also model the spatial characteristics as long-term stationary. All other experiment settings remain unchanged as described before in Section \ref{sec:4-b-inh-noise-speakers}. 

\subsubsection{Performance evaluation}
The performance results are displayed in the second column of Table \ref{table:multisource-results}. For this artificial type of noise, we observe an even greater performance difference of 9.9 dB \ac{SI-SDR} and 2.6 \ac{POLQA} score. In fact, the $\T{MMSE}$ estimator seems to be able to recover the original signal almost perfectly except from minor residual high-frequency noise while $\T{MVDR-MMSE}$ suffers from clearly audible speech degradation and residual noise. Audio examples can be found online\footref{link:audioexamples}.

Figure \ref{fig:4-b-specgrograms} depicts the spectrograms of the clean and noisy signals in the top row and the spectrograms of the enhancement results obtained by the $\T{MVDR}$, $\T{MVDR-MMSE}$, and $\T{MMSE}$ estimators in the bottom row. The uniform green coloration of the vertical stripes in the noisy spectrogram reflects the spectral stationarity. The vertical dark blue lines separate segments with different spatial properties. While the spatial diversity cannot be seen from the spectrogram, it becomes visible from the result of the \ac{MVDR} beamformer (first in bottom row). Here, the \ac{MVDR} beamformer suppresses different frequencies for signal segments with different spatial properties as can be seen from the displaced horizontal dark blue lines. The described differences between the $\T{MVDR-MMSE}$ (middle) and $\T{MMSE}$ (right) estimators' results are also found in the spectrograms. A close look reveals that the nonlinear spatial filter preserves much more of the target signal's fine structure. Furthermore, a comparison with the spectrogram of the clean speech signal highlights that it suppresses background noise much better than the $\T{MVDR-MMSE}$ estimator. Residual noise is visible in the spectrogram only in some segments at a frequency of about 6 kHz.

\subsubsection{Discussion}
\begin{figure}
    \centering
    \includegraphics{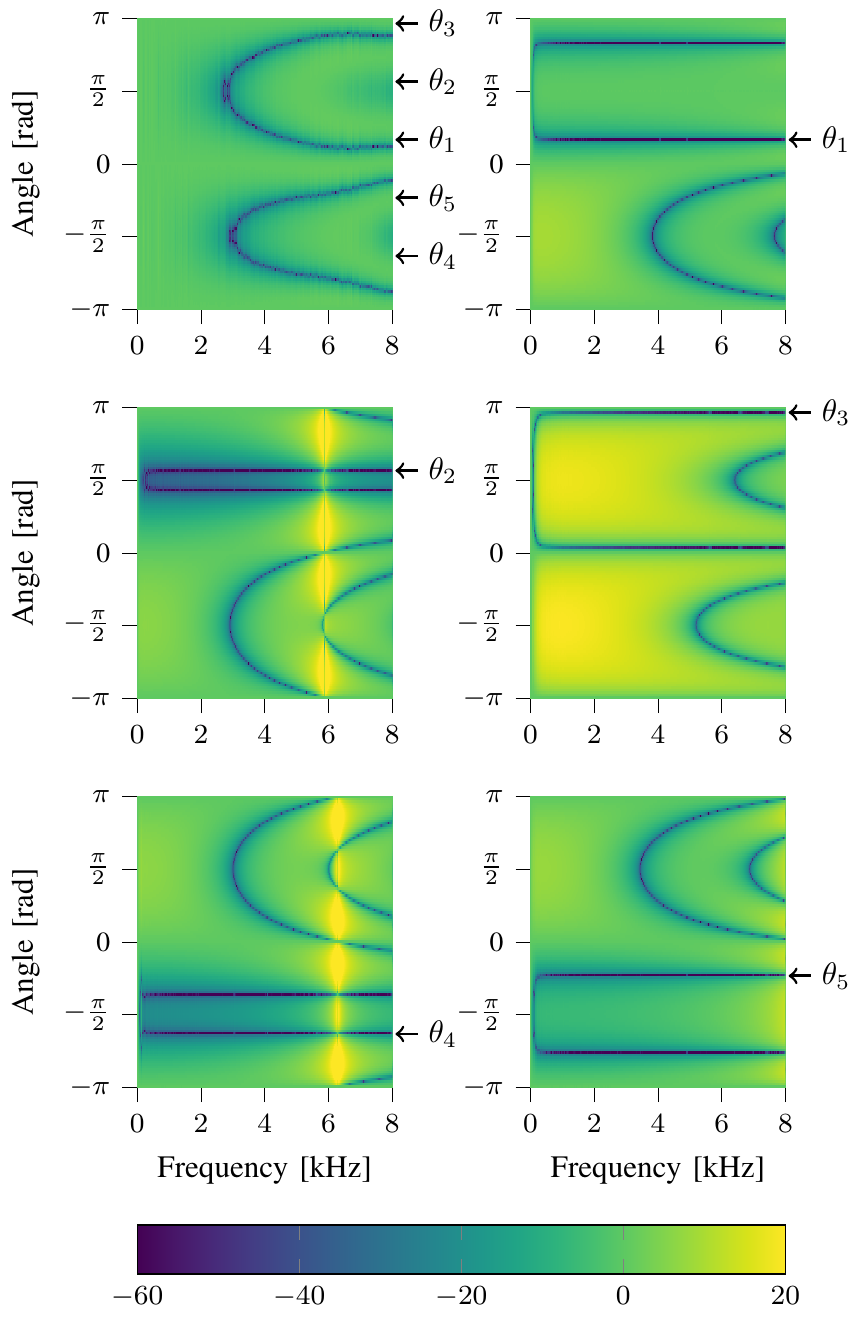}
    \caption{Directivity patterns of $\T{MVDR}$ (top left) and $T_{\text{MVDR}}^{(m)}$ (\ac{MVDR} beamformer with noise correlation matrix $\corM$), $m\in\{1,...,5\}$. The arrows on the right side indicate the incidence angle $\theta_i$, $i\in\{0,...,4\}$ of the $i$th interfering point source.}
    \label{fig:4-b-beampatterns}
\end{figure}

To explain these observations, we examine the covariance matrices $\corM$, $m\in\{1,...,5\}$, of the Gaussian mixture noise distribution estimated with the \ac{EM} algorithm. In Figure \ref{fig:4-b-beampatterns} we visualize their spatial structure based on the directivity pattern \cite[Sec. 12.5.2]{vary2006digital} that they produce when used as noise correlation matrix in the \ac{MVDR} beamformer, which is denoted with $T_{\text{MVDR}}^{(m)}$ in (\ref{eq:mmsegmm}). Furthermore, we visualize the directivity pattern of the \ac{MVDR} beamformer. The correlation matrix $\corN$ required to compute $\T{MVDR}$ is related to the mixture component covariance matrices via
\begin{IEEEeqnarray}{rCl}\label{eq:corrfromcomp}
\corN &=& \sum_{m=1}^M c_m\corM.
\end{IEEEeqnarray}
The directivity pattern produced by $\T{MVDR}$ is displayed at the top left followed by visualizations of the five mixture component covariance matrices. For each of these, a pronounced spatial characteristic can be observed by means of the horizontal dark lines. On the right side of the directivity patterns, we indicate the incidence angles $\theta_i$ of the noise sources. We notice that each component's covariance matrix models one of the noise sources as apparent from the zero placed in the respective direction by the \ac{MVDR} beamformer. The second horizontal line originates from the symmetry requirements of the directivity pattern, which are determined by the array geometry. 

In comparison, the directivity pattern of the $\T{MVDR}$ (top left) does not show zeros placed into the directions of the interfering point sources. Instead, the weighted combination in (\ref{eq:corrfromcomp}) seems to eliminate some of the spatial information, which corresponds to the well-known fact that a two-microphone \ac{MVDR} beamformer can suppress only one directional interfering source but not five of them. As a result, only some frequencies are suppressed for each interfering source as we observed in Figure \ref{fig:4-b-specgrograms} before. 

Figure \ref{fig:4-b-beampatterns} reveals how much more spatial information can be utilized by $\T{MMSE}$ in comparison with $\T{MVDR-MMSE}$, whose spatial processing relies on an estimate of $\corN$ as  visualized in the top left plot. The initial spatial filtering step using the \ac{MVDR} beamformer is not capable of suppressing the directional sources and the remaining noise has to be filtered by the postfilter, which then leads to some improvements. On the other hand, the nonlinear spatial $\T{MMSE}$ estimator can utilize the spatial information provided by the estimates of covariance matrices $\corM$. 

One could argue that a time-varying \ac{MVDR} beamformer $T_{\text{MVDR}}^{(m)}$ with correctly chosen $m$ would suffice to solve the problem on the short signal segments with noise from a single point source and a complicated nonlinear approach is not required. However, we must point out that the step of choosing the `right` covariance matrix is not required for the nonlinear spatial filter. Instead, we provide the Gaussian mixture parameters reflecting the spatial properties of the full utterance and, nevertheless, the $\T{MMSE}$ estimator is capable of suppressing five directional noise sources with only two microphones without the need for spatial adaptation. Note that this is an exciting finding as traditional linear spatial filters can only suppress $D-1$ interfering point sources with $D$ microphones without spatial adaptation \cite[Sec. 6.3]{trees2004OptimumArrayProcessing}.

Despite the impressive performance results achieved in this experiment, the analytic nonlinear spatial filter has some weaknesses: it requires a very accurate estimation of the spatial and spectral characteristics of the noise signal and is also computationally quite demanding. In addition to the presented experiments, we carried out simulations using measured impulse responses between the microphones and speakers and observed a much lower benefit from using a nonlinear spatial filter for the experiment with five interfering Gaussian sources even with access to oracle noise data. 
This is because the spatial and spectral diversity of the noise signal increase and many more mixture components would be required to model the noise accurately which then results in a data problem. Similarly, estimating the parameters of the Gaussian mixture from a noisy signal is difficult. We approached this using masks to identify time-frequency bins that are dominated by noise but did not obtain reliable estimates this way.

Therefore, we conclude that the analytical estimators allow us to study the potential of nonlinear spatial filters in principle, but because of high sensitivity to parameter estimation errors and high computational costs, practical nonlinear spatial filters may be better implemented using modern machine learning tools like \acp{DNN}.

\section{Conclusions}
In a detailed theory overview, we have revisited the fact that the multichannel \ac{MMSE}-optimal estimator of the clean speech signal is in general a jointly spatial-spectral nonlinear filter. Therefore, the state-of-the-art concatenation of a linear spatial filter and a postfilter is \ac{MMSE}-optimal only in the special case that the noise follows a Gaussian distribution. The experimental section of this paper studied the performance advantage that can be gained by replacing the generally suboptimal sequential setup with a nonlinear spatial filter in three different non-Gaussian noise scenarios. 

First, we  have  shown  that  considerable performance improvements  result  if  the  noise  distribution  deviates from a Gaussian distribution by an increased heavy-tailedness  as the nonlinear spatial filter enables a higher noise reduction and lower speech distortions at the same time. Second, we report a performance benefit of 4.6 dB \ac{SI-SDR} and of 0.8 \ac{POLQA} score for an inhomogeneous noise field created by five interfering speech sources and, furthermore, we have observed a benefit of about 3.2 dB \ac{SI-SDR} and 0.6 POLQA score for the real-world cafeteria noise recordings from the CHiME3 database. In addition, we have performed experiments that revealed that the nonlinear spatial filter has some notably increased spatial processing capabilities allowing for an almost perfect elimination of five Gaussian interfering point sources with only two microphones. 

The presented findings on the performance potential of a nonlinear spatial filter motivate further research on the implementation of nonlinear spatial filters, e.g., using \acp{DNN} to learn the nonlinear spatial filter directly from data and overcome the parameter estimation issues and other limitations of the analytic nonlinear spatial filter that we have used for this analysis.

\section*{Acknowledgment}
We would like to thank J. Berger and Rohde\&Schwarz SwissQual AG for their support with POLQA.

\bibliographystyle{IEEEtran}
\bibliography{bib}

\begin{IEEEbiography}[{\includegraphics[width=1in,height=1.25in,clip,keepaspectratio]{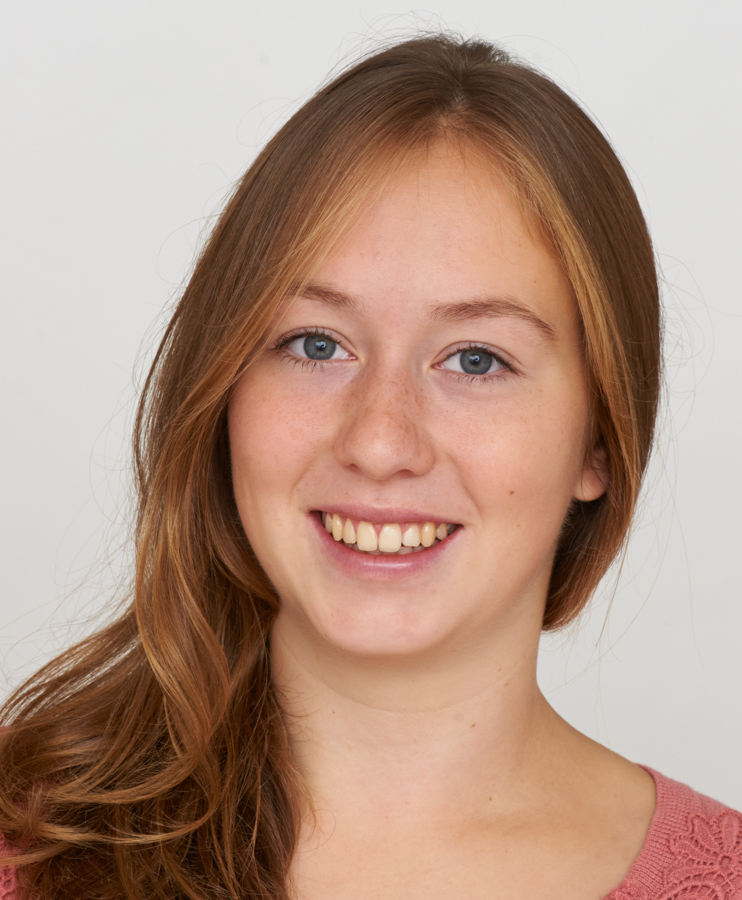}}]{Kristina Tesch}
(S’20) received a B.Sc and M.Sc in Informatics in 2016 and 2019 from the Universität Hamburg, Hamburg, Germany. She is with the Signal Processing Group at the Universität Hamburg since 2019 and is currently working towards a doctoral degree. Her research interests include digital signal processing and machine learning algorithms for speech and audio with a focus on multichannel speech enhancement. For her master thesis on multichannel speech enhancement, she received the award for the best master thesis at a German Informatics Department from the Fakultätentag Informatik in 2019.
\end{IEEEbiography}

\begin{IEEEbiography}[{\includegraphics[width=1in,height=1.25in,clip,keepaspectratio]{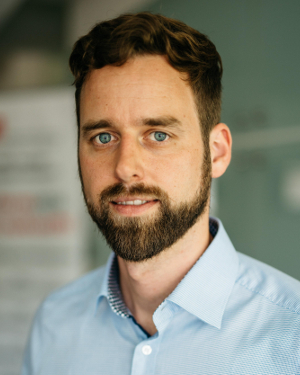}}]{Timo Gerkmann}
(S’08–M’10–SM’15) studied Electrical Engineering and Information Sciences at the Universität Bremen and the Ruhr-Universität Bochum in Germany. He received his Dipl.-Ing. degree in 2004 and his Dr.-Ing. degree in 2010 both in Electrical Engineering and Information Sciences from the Ruhr-Universität Bochum, Bochum, Germany. In 2005, he spent six months with Siemens Corporate Research in Princeton, NJ, USA. During 2010 to 2011 Dr. Gerkmann was a postdoctoral researcher at the Sound and Image Processing Lab at the Royal Institute of Technology (KTH), Stockholm, Sweden. From 2011 to 2015 he was a professor for Speech Signal Processing at the Universität Oldenburg, Oldenburg, Germany. During 2015 to 2016 he was a Principal Scientist for Audio \& Acoustics at Technicolor Research \& Innovation in Hanover, Germany. Since 2016 he is a professor for Signal Processing at the Universität Hamburg, Germany. His main research interests are on statistical signal processing and machine learning for speech and audio applied to communication devices, hearing instruments, audio-visual media, and human-machine interfaces. Timo Gerkmann serves as an elected member of the IEEE Signal Processing Society Technical Committee on Audio and Acoustic Signal Processing and as an Associate Editor of the IEEE/ACM Transactions on Audio, Speech and Language Processing.
\end{IEEEbiography}

\end{document}